\documentclass[final]{aipproc}
\layoutstyle{8x11single}
\newcommand{\inieq}{\begin{eqnarray}}            
\newcommand{\fineq}{\end{eqnarray}}            
\newcommand{\diff}{{\rm\,d}}                    
\newcommand{\be}{\begin{equation}}
\newcommand{\ba}{\begin{eqnarray}}
\newcommand{\ea}{\end{eqnarray}}

\def\ee{\mbox{$\left(e,e^{\prime}\right)$\ }}
\def\eep{\mbox{$\left(e,e^{\prime}p\right)$\ }}

\def\mcv{\mbox{$\mathcal{V}$}}
\begin{document} 
\title{Quasielastic scattering with
the Relativistic Green's Function approach}



\author{Andrea Meucci}{
  address={Dipartimento di Fisica  - Universit\`a degli Studi di Pavia\\ and INFN - Sezione di Pavia, 
  via A. Bassi 6, I-27100 Pavia, Italy}
}

\author{Carlotta Giusti}{
  address={Dipartimento di Fisica  - Universit\`a degli Studi di Pavia\\ and INFN - Sezione di Pavia, 
  via A. Bassi 6, I-27100 Pavia, Italy}
}

\begin{abstract}
A relativistic model for quasielastic (QE) lepton-nucleus scattering is  presented.
The effects of
final-state interactions (FSI) between the ejected nucleon
and the residual nucleus are described in the
relativistic Green's function (RGF) model where  FSI are consistently
described with exclusive 
scattering using a complex optical potential. The results of the model are compared with 
experimental results of electron
and neutrino scattering.
\end{abstract}

\maketitle

\section{Final-state interactions in  lepton-nucleus scattering}

In the QE region the nuclear response to an electroweak probe is dominated by
one-nucleon knockout processes, where the scattering occurs
with only one nucleon while the
remaining nucleons of the target behave as simple spectators. The 
reaction can adequately be described in
the relativistic impulse approximation (IA) by the sum of
incoherent processes involving only one nucleon scattering and the components 
of the hadron tensor  are obtained from the 
sum, over all the single-particle (s.p.) shell-model states, of the squared absolute value 
of the transition 
matrix elements of the single-nucleon current.
A reliable description of FSI  is an essential ingredient for the
comparison with data.
The relevance of FSI has been clearly stated for the exclusive 
$\eep$ reaction, where the use of 
complex optical potentials (OP) in the distorted-wave impulse approximation (DWIA) 
is required~\cite{Boffi:1993gs,book,Udias:1993xy,Meucci:2001qc,
Meucci:2001ja,Meucci:2001ty,Radici:2003zz,Giusti:2011it}. 
 The imaginary part of the OP produces an absorption that reduces
the cross section and accounts
for the fact that, if other channels
are open besides the elastic one, part of the incident flux is
lost in the elastically scattered beam and goes to the other
(inelastic) channels which are open.
In the inclusive scattering only the emitted lepton is detected,
the final nuclear state is not determined and  all
elastic and inelastic channels contribute. This requires a
different treatment of FSI where
all final-state channels should be retained
and the total flux, although redistributed among all possible
channels, must be conserved.
Different
approaches have been used to describe FSI in relativistic
calculations for the inclusive QE electron- and neutrino-nucleus
scattering. In the relativistic plane-wave
impulse approximation (RPWIA), FSI are simply neglected.
In another\ approach, FSI are accounted for in relativistic DWIA (RDWIA)
calculations by including only the real part of the relativistic optical
potential (rROP). 

In the RGF 
techniques~\cite{Capuzzi:1991qd,Meucci:2003uy,Meucci:2003cv,Capuzzi:2004au,Meucci:2005pk,
Meucci:2006cx,Giusti:2009sy,
Meucci:2009nm,Meucci:2011pi,Meucci:2011vd,Meucci:ant}, FSI are 
described in the inclusive scattering by the 
same complex OP as in the exclusive scattering, but the imaginary part is used in the 
two cases in a different way and in the inclusive reaction the flux, although is redistributed 
in all the channels, is conserved.
In the RGF model with suitable approximations, which are mainly
related to the impulse approximation, the components of the hadron tensor are 
written in terms of the s.p. optical model Green\rq{}s
function. The explicit calculation of the s.p. Green's function can be 
avoided by its spectral representation, which is based
on a biorthogonal expansion in terms of the eigenfunctions of the non-Hermitian 
optical potential and of its Hermitian conjugate
\inieq { 
\left[ {\mathcal{E}} - T - {\mathcal{V}}^{\dagger} (E) \right] \mid
{\chi}_{\mathcal{E}}^{(-)}(E)\rangle = 0\ , \ \ \ 
\left[ \mathcal{E} - T - {\mathcal{V}}(E) \right] \mid \tilde
{\chi}_{\mathcal{E}}^{(-)}(E)\rangle = 0\ } \ .
\fineq
The expanded form for the s.p. expression of the hadron tensor 
components is \cite{Meucci:2003uy,Meucci:2003cv}
\inieq
W^{\mu\nu}(q,\omega)  =  \sum_n \Bigg[ \mathsf{Re} \ T_n^{\mu\nu}
(E_{{f}}-\varepsilon_n, E_{{f}}-\varepsilon_n)  
~ ~ 
- \frac{1}{\pi} \mathcal{P}  \int_M^{\infty} \diff \mathcal{E} 
\frac{1}{E_{{f}}-\varepsilon_n-\mathcal{E}} 
\ \mathsf{Im} \ T_n^{\mu\nu}
(\mathcal{E},E_{{f}}-\varepsilon_n) \Bigg] \ , \label{eq.finale}
\fineq
where $\mathcal{P}$ denotes the principal value of the integral, $n$ is the 
eigenstate of the residual nucleus with energy 
$\varepsilon_n$, and
\inieq
T_n^{\mu\mu}(\mathcal{E} ,E) = \lambda_n  \langle \varphi_n
\mid j^{\mu\dagger}(\mathbf{{q}}) \sqrt{1-\mcv'(E)}
\mid\tilde{\chi}_{\mathcal{E}}^{(-)}(E)\rangle 
~    \langle\chi_{\mathcal{E}}^{(-)}(E)\mid  \sqrt{1-\mcv'(E)} j^{\mu}
(\mathbf{q})\mid \varphi_n \rangle  \ , \label{eq.tprac}
\fineq
and similar\ expressions for the terms with $\mu \neq \nu$.
The factor $\sqrt{1-\mcv'(E)}$, where $\mcv'(E)$ is the energy derivative of 
the optical potential, accounts for interference effects between different 
channels and  allows the replacement of
the mean field ${\mcv}$ with the phenomenological OP.

Disregarding the square root correction, the second matrix element in 
Eq.~(\ref{eq.tprac}) is the transition amplitude for 
the single-nucleon knockout 
from a nucleus in the state { $\mid \Psi_0\rangle$} leaving the 
residual nucleus 
in the state { $\mid n \rangle$} and it
is similar to the usual DWIA expression for the transition amplitude of the exclusive 
single-nucleon knockout, i.e., the imaginary part of { $\mcv^{\dagger}$}   gives 
an attenuation of the strength 
that 
is inconsistent with the inclusive process, where all the inelastic channels 
must be considered and the total flux must be conserved.
This compensation is performed by the first matrix element in the right hand 
side of Eq.~(\ref{eq.tprac}),  which involves the eigenfunction 
$\tilde{\chi}_{\mathcal{E}}^{(-)}(E)$ of the Hermitian conjugate optical
potential, where the imaginary part has an opposite sign and has the 
effect of increasing the strength. Therefore, in the RGF approach the 
imaginary part of the optical
potential redistributes the flux lost in a channel in the other channels, and 
in the sum over $n$ the total flux is conserved.  The
RGF model
 requires the calculations of matrix elements of the same type as in usual
 RDWIA models, but involves eigenfunctions of both { $\mcv$} and
{ $\mcv^{\dagger}$}: FSI are described by the same complex optical potential as in RDWIA
thus providing a consistent treatment of FSI in the exclusive and in the
inclusive scattering.

\section{Results for electron and neutrino scattering}

The first measurement of the charged-current quasielastic (CCQE)
flux-averaged double-differential muon neutrino cross
section on $^{12}$C in an energy range up to $\approx$ 3 GeV that
have  been reported by the MiniBooNE 
Collaboration  \cite{miniboone} have raised  extensive
discussions.
In particular, the experimental cross section is usually 
 underestimated by the
relativistic Fermi gas  model and by other more sophisticated models based
on the IA \cite{Benhar:2010nx,Butkevich:2010cr,jusz10}, unless the nucleon axial mass  
$M_A$ is significantly enlarged with respect to the world average value of 1.03 GeV/$c^2$. 
Despite the fact the larger axial mass obtained from the MiniBooNE data on 
carbon can  be interpreted as an effective way to include medium effects 
which are not taken into account, it is clear that, 
before drawing conclusions,  a precise knowledge of lepton-nucleus cross 
sections, where uncertainties on nuclear effects are reduced as much as 
possible, is necessary. 
Moreover, any reliable calculation for
neutrino scattering should first be tested against electron
scattering in the same kinematical conditions.

In this section several results obtained with the RGF model for electron 
and neutrino-nucleus scattering are discussed. 
Some results of the RGF  for  inclusive electron and neutrino
scattering are presented in \cite{Meucci:2009nm,Meucci:2011vd}, where they are 
compared with the results
obtained with the relativistic mean
field  model \cite{PhysRevLett.95.252502}, which uses the same strong real 
potential already considered in describing the bound states to evaluate 
the scattering wave functions.
\begin{figure}[t]
\includegraphics[scale=.33]{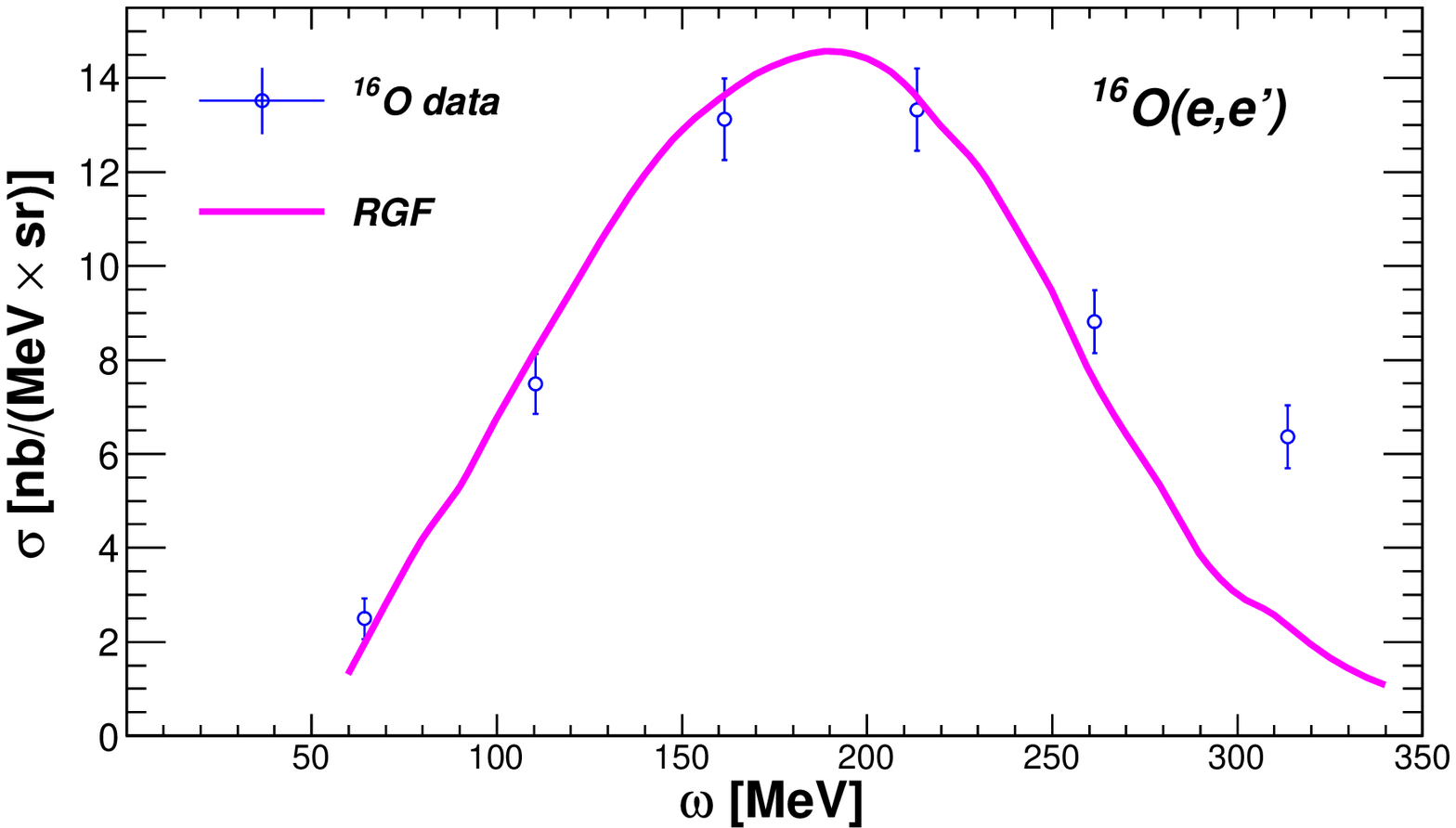}
\includegraphics[scale=.33]{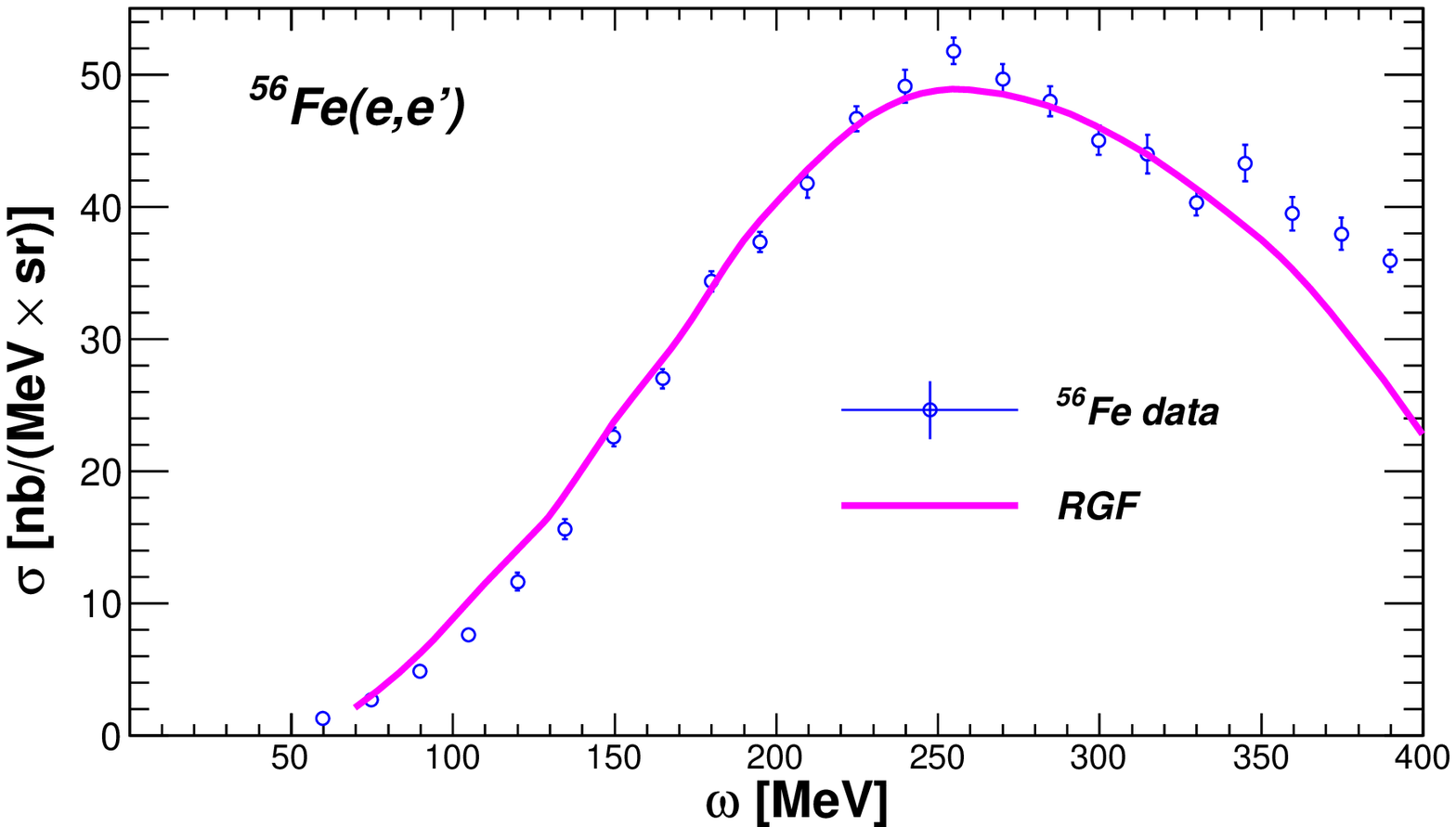}
\caption{
Differential cross section of the reactions
$^{16}$O$(e,e^{\prime})$  for  beam energy  
$\varepsilon = 1080$ MeV and scattering angle $\vartheta = 32^{\mathrm{o}}$
 and	$^{56}$Fe$(e,e^{\prime})$  for  beam energy  
$\varepsilon = 2020$ MeV and scattering angle $\vartheta = 20^{\mathrm{o}}$. 
Experimental data from \cite{Anghinolfi:1996vm} ($^{16}$O) 
and \cite{fe56} ($^{56}$Fe).}\label{fexpgf}
	\end{figure}

As an example, in Fig.~\ref{fexpgf} the RGF results are compared with the 
experimental $\ee$ cross section on $^{16}$O and $^{56}$Fe. 
In all the calculations presented here the bound nucleon states 
are self-consistent Dirac-Hartree solutions derived within a relativistic
mean-field approach~\cite{Serot:1984ey}, and
in the RGF different parameterizations have been used for the relativistic
optical potential: the 
energy-dependent and A-dependent (where A is the mass number) EDAD1 and the 
energy-dependent but A-independent EDAI complex phenomenological potentials 
of~\cite{Cooper:2009}.
The shape followed by the RGF cross sections fits well the slope shown 
by the data,  in
particular approaching the peak region, where  the RGF produces cross sections 
in reasonable agreement with the data. 
Although satisfactory on general grounds, the comparison with data in 
Fig.~\ref{fexpgf} 
cannot be conclusive until contributions beyond the QE peak, like meson 
exchange currents and $\Delta$ effects, which may
play a significant role in the analysis of data even at the maximum of the 
QE peak, are carefully evaluated.

	\begin{figure}[tb]
\includegraphics[scale=.25]{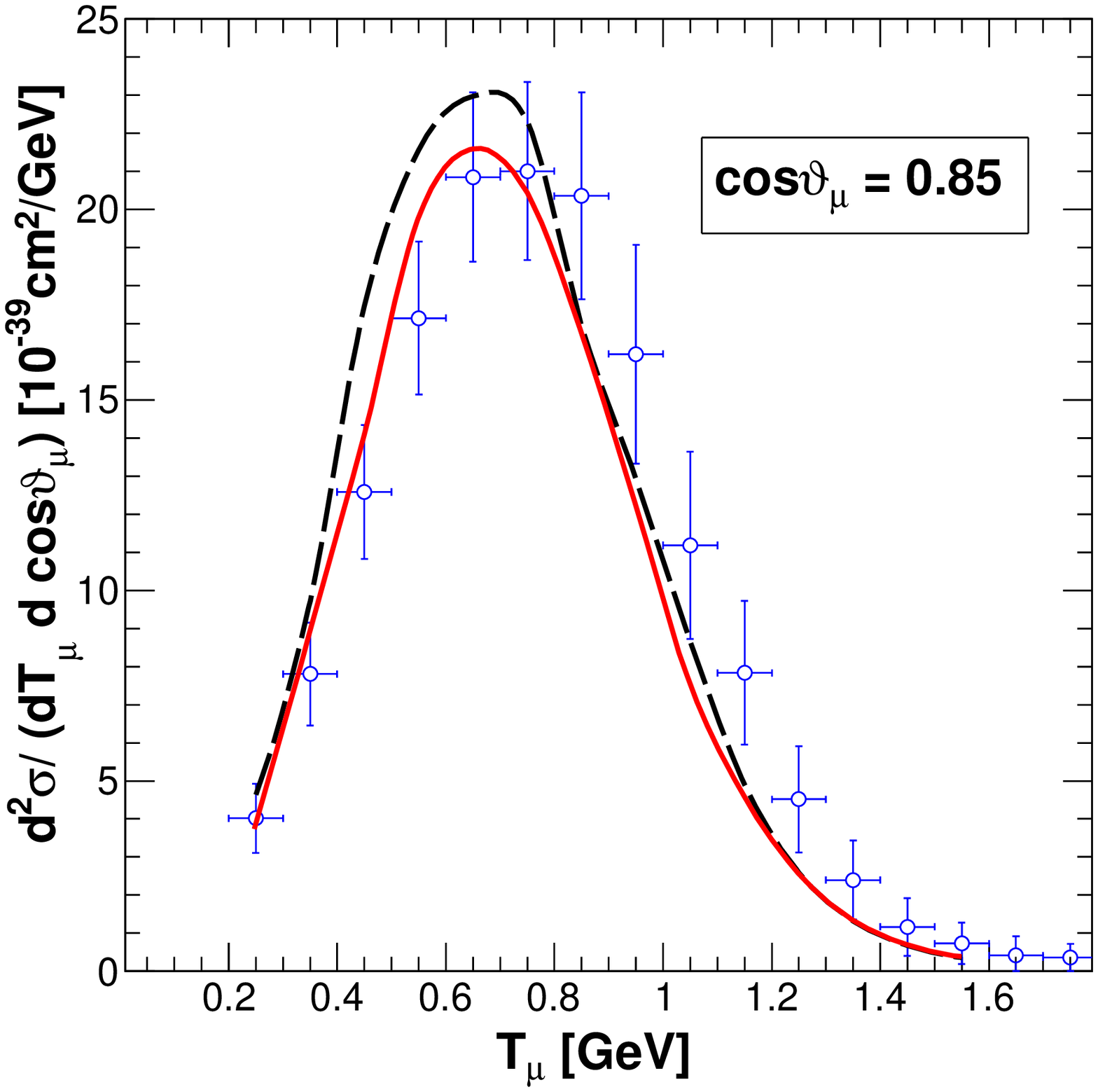}
\includegraphics[scale=.25]{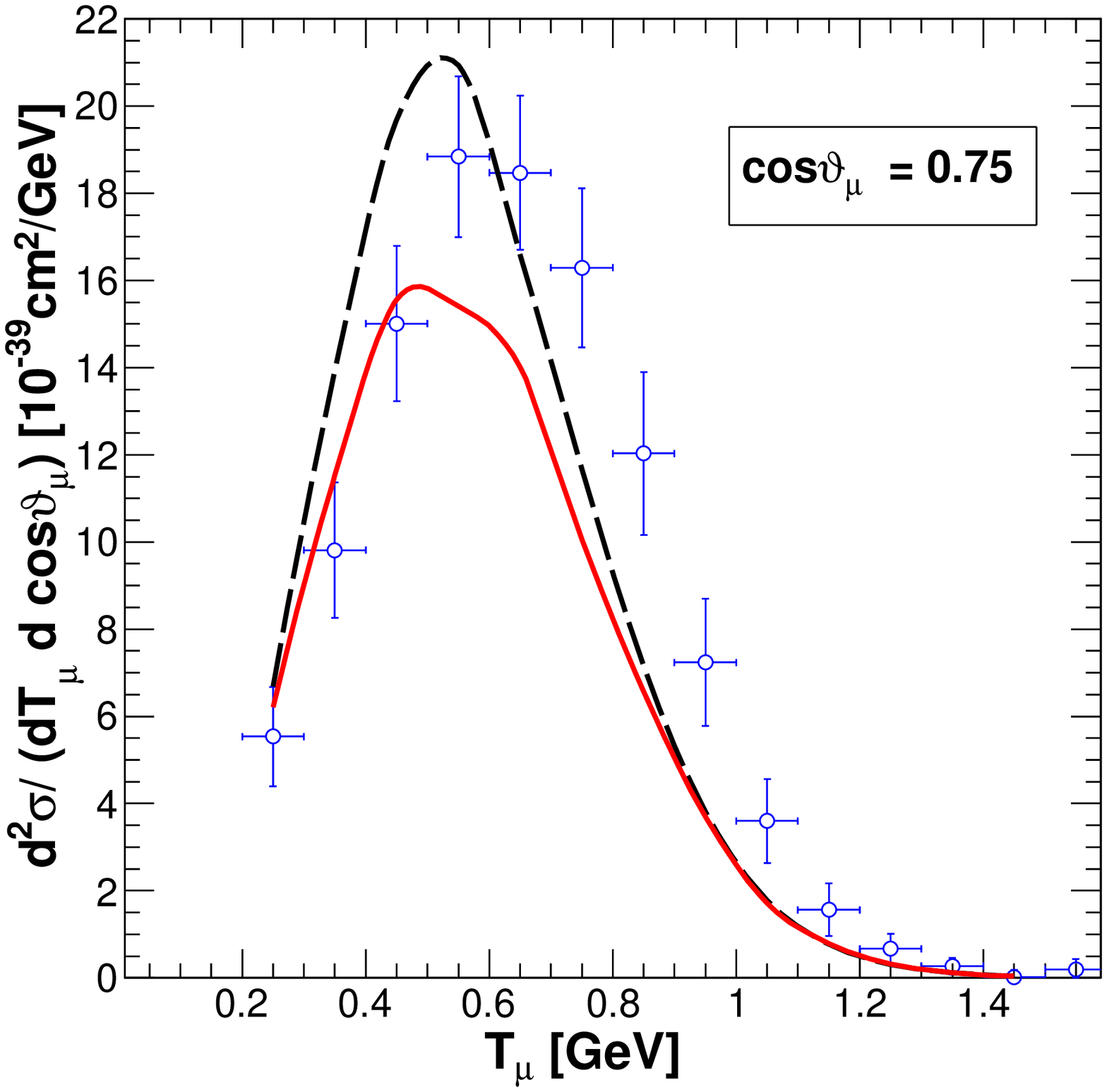}
\includegraphics[scale=.25]{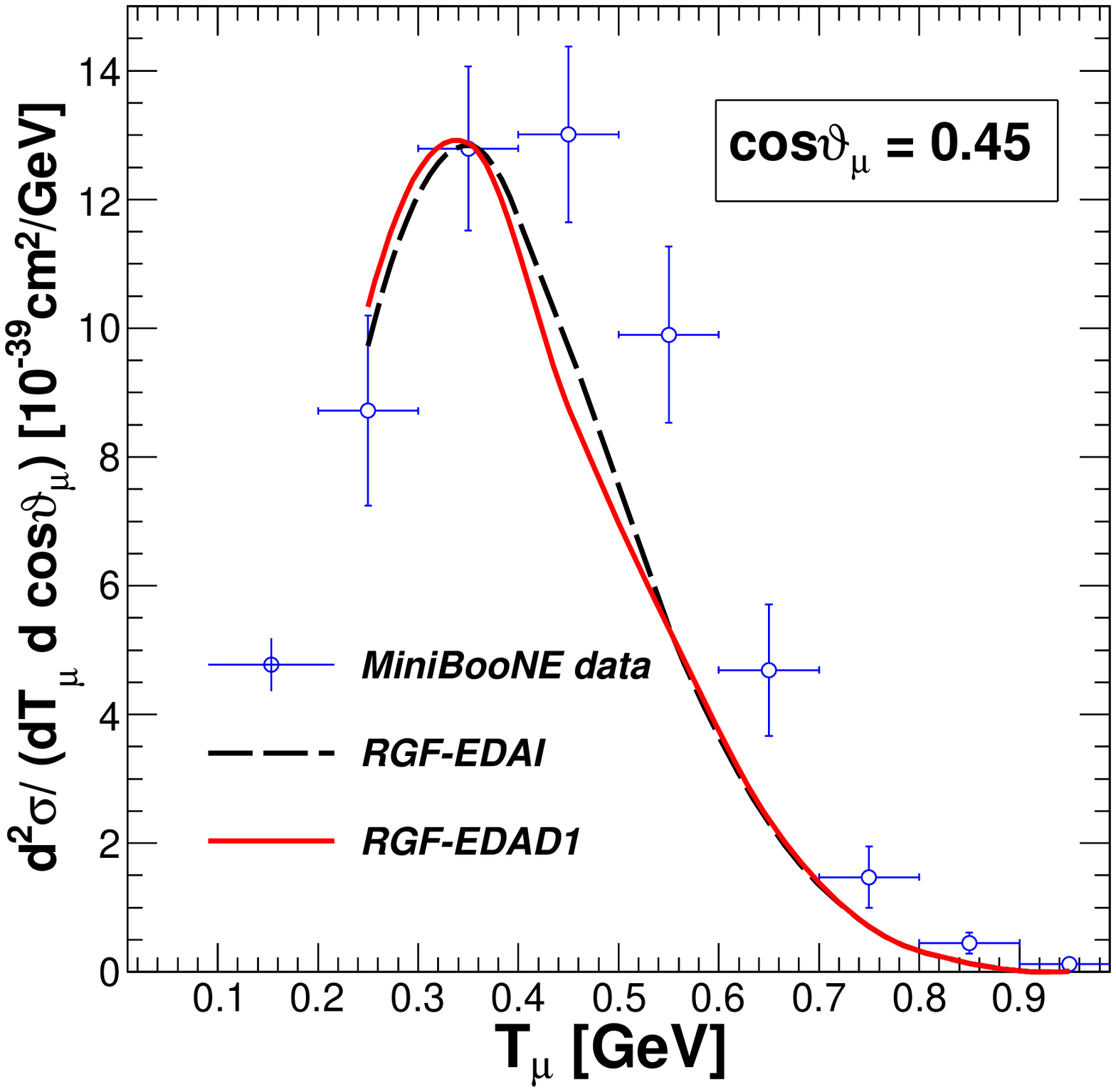}
\caption{Flux-averaged double-differential cross section per 
target nucleon for the CCQE $^{12}$C$(\nu_{\mu} , \mu ^-)$ reaction calculated 
with the  
	RGF-EDAI (dashed lines) and the RGF-EDAD1 (solid lines) displayed versus $T_\mu$ for three bins 
	of $\cos\vartheta_{\mu}$. 
The data are from MiniBooNE \cite{miniboone}.
}\label{neu-double}
	\end{figure}

In Fig.~\ref{neu-double} the CCQE double-differential 
$^{12}$C$(\nu_{\mu} , \mu ^-)$  cross sections
averaged over the  neutrino flux is displayed as a function of the muon kinetic energy 
$T_{\mu}$ for three bins of $\cos\vartheta_{\mu}$, where $\vartheta_{\mu}$ is the muon 
scattering angle. 
In all the calculations of neutrino-nucleus scattering the standard value of 
the nucleon axial mass,  i.e., $M_A = 1.03$ GeV/$c^2$ has been used.
A good agreement with the MiniBooNE data of \cite{miniboone}
is generally shown  by the RGF cross sections \cite{Meucci:2011vd}. 
The differences between the RGF
results with the two optical potentials are enhanced
in the peak region but they always are of the order
of the experimental errors. The EDAD1 and EDAI potentials
yield close predictions for the bin $0.4 \leq \cos\vartheta_{\mu} \leq 0.5$;
small differences are seen in the bin $0.8 \leq \cos\vartheta_{\mu} \leq 0.9$, while 
larger differences can be found for the bin
$0.7 \leq \cos\vartheta_{\mu} \leq 0.8$. Nevertheless, the
RGF-EDAI cross section is always larger than the RGF-EDAD1
one. The differences between the RGF-EDAI and the 
RGF-EDAD1 results are due to the different imaginary parts. 

	\begin{figure}
\includegraphics[scale=0.3]{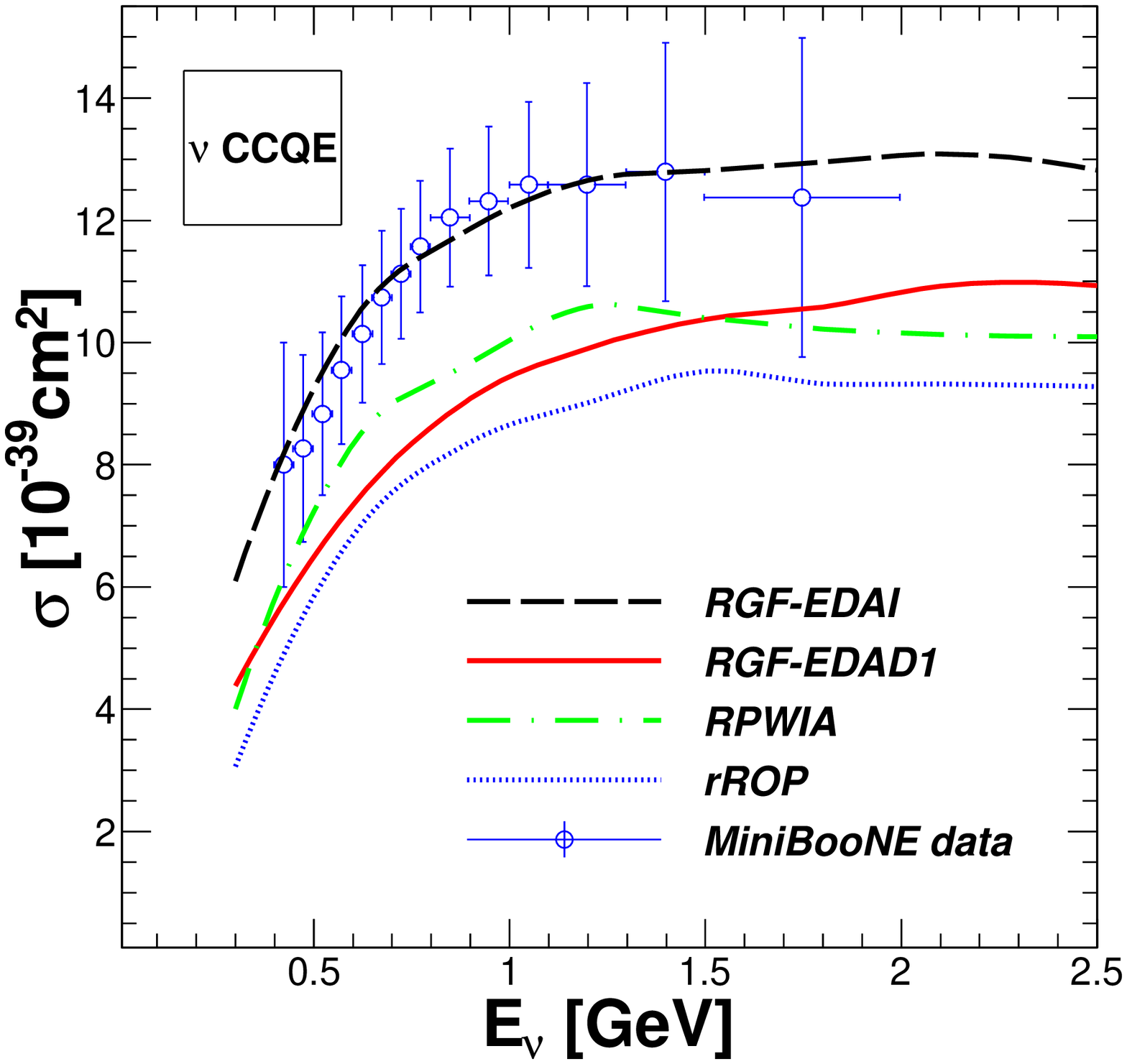}
\includegraphics[scale=0.3]{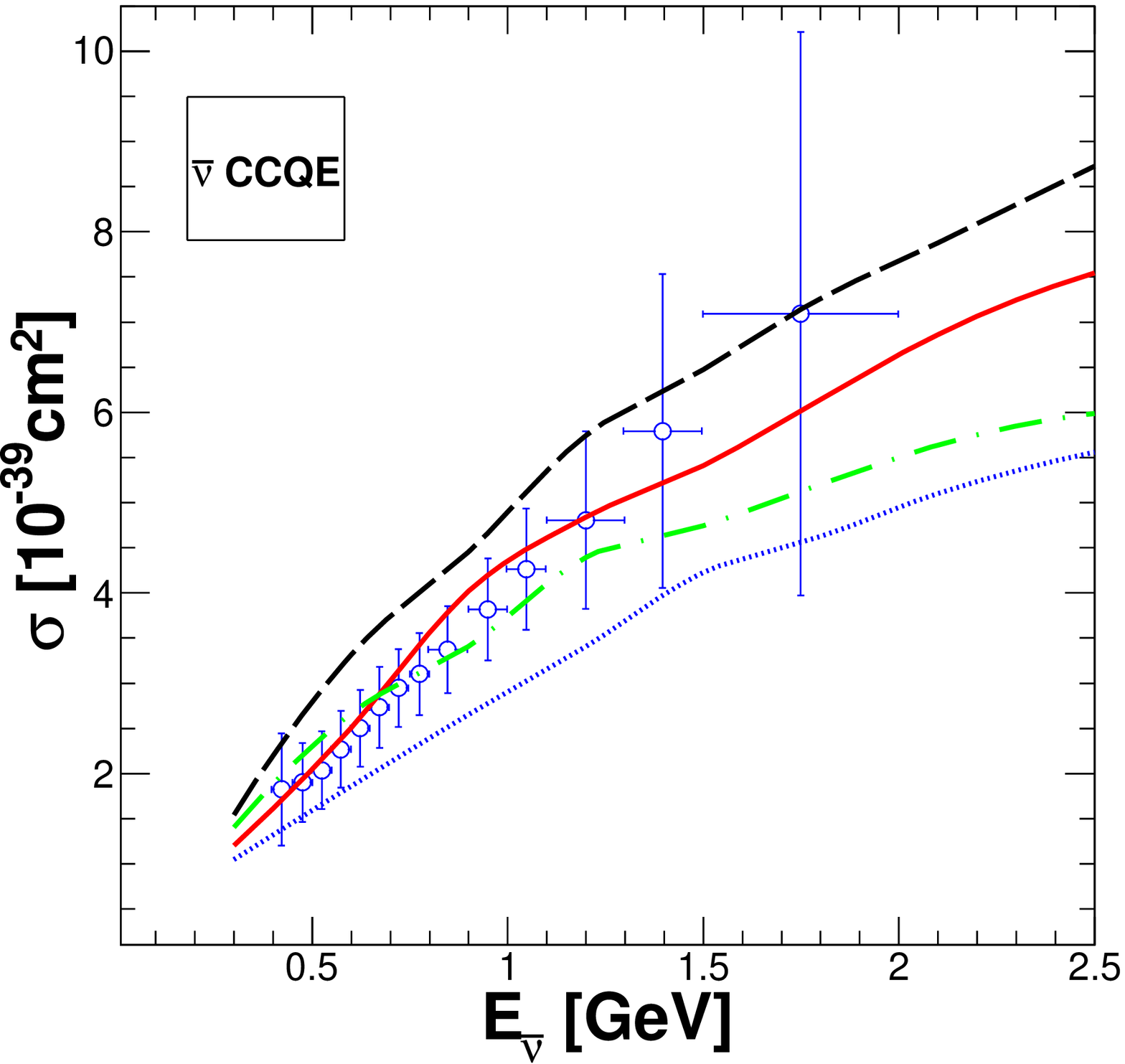}
\caption{Total CCQE cross section per target nucleon as a function of 
$E_{\nu}$ (left panel) and $E_{\bar\nu}$ (right panel) 
calculated with the RGF-EDAD1 (solid line), the 
RGF-EDAI (dashed line),  the rROP 
(dotted line), and the RPWIA (dot-dashed line). 
The data are from MiniBooNE
\cite{miniboone,AguilarArevalo:2013hm}.
}\label{figcc}
\end{figure}

In Fig.~\ref{figcc} the total CCQE cross sections per target nucleon for
neutrino and antineutrino scattering
are displayed as  functions of the neutrino or antineutrino energies 
$E_{\nu}$ and $E_{\bar\nu}$ and
compared with the \lq\lq unfolded\rq\rq\ MiniBooNE 
data \cite{miniboone,AguilarArevalo:2013hm}. 
The RPWIA and rROP results usually underpredict
the $\nu_{\mu} $ data. Larger cross sections, in particular for larger
values of $E_{\nu}$, are obtained in the RGF with both optical
potentials \cite{Meucci:2011vd}. The differences between the RGF-EDAI
and the RGF-EDAD1 results, being RGF-EDAI in good
agreement with the shape and magnitude of the $\nu_{\mu}$
data, are due to the different imaginary
parts. The enhancement of the RGF cross
sections is due to the translation to the inclusive strength
of the overall effect of inelastic channels which are not
incorporated  in other models based on the IA.

The total CCQE cross section for $\bar\nu_{\mu} $ scattering 
is displayed in the right panel of Fig.~\ref{figcc}. 
Also in this case the RGF results are usually larger than the RPWIA 
and rROP ones. The differences between the 
RGF-EDAD1 and RGF-EDAI results are significant but somewhat smaller than for 
neutrino scattering. A 
reasonable agreement with the experimental data is obtained, but the 
RGF-EDAI calculations are larger than the data
up to $E_{\bar\nu} \approx 1.5$ GeV, while a better agreement is obtained with the RGF-EDAD1 ones.
The different behavior of the cross 
sections for neutrino and antineutrino scattering is related to the 
relative strength of the vector-axial response,  which is constructive in 
$\nu$ scattering and destructive in $\bar\nu$ scattering, with respect to the longitudinal and 
transverse ones \cite{Meucci:ant}.
	\begin{figure}
\includegraphics[scale=0.3]{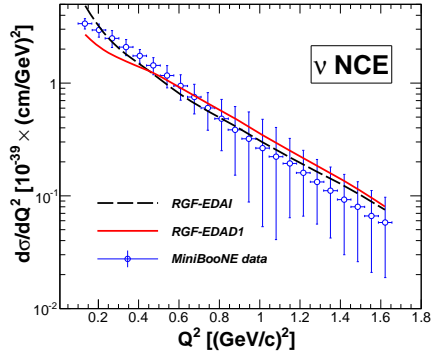}
\caption{NCE  flux-averaged cross section per target nucleon as a function of 
$Q^{2}$  
 calculated with the RGF-EDAD1 (solid line) and the 
RGF-EDAI (dashed line).
The data are from MiniBooNE
\cite{miniboonenc}.
}\label{fignc}
\end{figure}

The MiniBooNE Collaboration has reported \cite{miniboonenc}  also a
measurement of the flux-averaged differential cross section as a function of 
the four-momentum transferred squared, $Q^2 = -q^{\mu}q_{\mu}$, for 
neutral-current elastic (NCE) neutrino scattering on CH$_2$ in a $Q^2$ range up 
to $\approx 1.65\ ($GeV/$c)^2$.
The analysis of $\nu$-nucleus NCE reactions introduces additional 
complications, as the final neutrino cannot be measured and a 
final hadron has to be detected: the cross 
sections are therefore semi-inclusive in the hadronic sector and inclusive in 
the leptonic one \cite{Meucci:2004ip,Meucci:2006ir,Meucci:2008zz}. 
Different relativistic descriptions of FSI are  
compared with the NCE MiniBooNE data in \cite{Meucci:2011nc}, while
in Fig.~\ref{fignc} we show our RGF results calculated with both
EDAI and EDAD1 potentials.
Also in this case, the RGF produces large cross sections
that are in nice agreement with the data. However, we stress 
the RGF is appropriate for the inclusive scattering
where only the final lepton is detected, and thus can take
into account also contributions that are not included in the
experimental cross sections.

\section{Summary and conclusions}

A deep understanding of the 
neutrino-nucleus cross sections is very important for the determination
of neutrino oscillation parameters. Reliable
theoretical models are required where all nuclear effects
are well under control. Within the QE kinematics domain,
the treatment of FSI is an essential ingredient for
the comparison with data. In this contribution
the RGF model for the inclusive QE electron
and neutrino-nucleus scattering has been discussed.
This model was originally developed for QE
electron scattering, successfully tested in comparison
with electron-scattering data, and  then  extended
to neutrino-nucleus scattering.
In the RGF model FSI
are described in the inclusive scattering by the same complex
optical potential as in the exclusive scattering, but
the imaginary part is used in the two cases in a different
way and in the inclusive process it is responsible
for the redistribution of the flux in all the channels and
the conservation of the  flux. The RGF model gives results that 
are usually larger than results of other models based on the impulse
approximation and that are in fair agreement with the CCQE MiniBooNE 
cross sections without the need to increase the standard value of the
nucleon axial mass.
However, the use of phenomenological optical potentials,
does not allow us to disentangle the role of different
reaction processes and explain in detail the origin
of the enhancement with respect of other  models. 
The important role of contributions
other than direct one-nucleon emission has been
confirmed by different models; 
it has been observed that
the neutrino-nucleus reaction at MiniBooNE can have significant contributions
from effects beyond the IA in some kinematic
regions where the experimental neutrino flux has significant
strength. In Refs. \cite{Benhar:2010nx,Martini:2010ex,Martini:2011wp,Nieves:2011pp,
Nieves:2011yp,PhysRevC.86.014614} 
the contribution of multinucleon excitations to CCQE
scattering has been found sizable and able to bring the
theory in agreement with the  MiniBooNE
cross sections without increasing the value of $M_A$. A critical review
of nuclear effects in NCE and CCQE scattering is presented in 
\cite{PhysRevC.86.024616}.
Moreover, processes involving two-body currents, whose role is
 discussed in \cite{Amaro:2011qb},  should also be
taken into account explicitly and consistently in a model
to clarify the role of multinucleon emission.
Fully relativistic
microscopic calculations of two-particle-two-hole (2p-2h)
contributions are extremely difficult and may be bound to
model-dependent assumptions.
The RGF results are also affected by uncertainties in
the determination of the phenomenological optical potential.
At present, lacking a phenomenological OP which exactly fulfills 
the dispersion relations in the whole energy region of 
interest, the RGF prediction is not univocally determined from the 
elastic phenomenology. A better determination
of a phenomenological relativistic optical potential,
which closely fulfills the dispersion relations, would be
anyhow desirable and deserves further investigation.

\bibliographystyle{aipproc}

\end{document}